\documentclass[twocolumn,a4paper,amsmath,amssymb,showpacs]{revtex4}

\usepackage[dvips]{epsfig}
\usepackage{dcolumn}% Align table columns on decimal point
\usepackage{bm}% bold math

\def\A{{K}}
\def\a{\kappa}

\def\opket#1#2 {  {#1} |{#2}\rangle }

\def\C{{\Gamma}}

\def\d{{\rm d}}

\def\del{\gamma}
\def\ddel{\Phi}

\def\dydxv#1#2{ {{\partial #1} \over {\partial #2}} }

\def\e{{\rm e}}

\def\eps{\varepsilon}

\def\ha{{1 \over 2}}

\def\Ibar{\bar{I}}

\def\Im{{\rm Im}}

\def\K2{{\cal K}}

\def\Lbar{\bar{L}}

\def\M{{\cal M}}

\def\Re{{\rm Re}}

\def\v{{u}}
\def\V{U}

\def\x{{\bf x}}

\def\sbar{\big/}
\def\dbar{ {\,\sbar\!\sbar} }
\def\dbarr{ {\,\sbar} }

\def\opsket#1#2#3 {  {#1} \dbar_{#2}{#3}\rangle }

\def\opssket#1#2#3 {  {#1} \dbarr_{#2}{#3}\rangle }

\begin{document}
\bibliographystyle{unsrt}
\title{Directional emission from weakly eccentric resonators}
\author{Stephen C. Creagh}
%\email{stephen.creagh@nottingham.ac.uk}
\affiliation{School of Mathematical Sciences,
University of Nottingham, University Park, Nottingham, NG7~2RD, UK}
\pacs{03.65.Sq,03.65.Xp,05.45.Mt,42.15.Dp,42.60.Da,42.65.Sf}
\begin{abstract}
It is shown that when a circular resonator is
deformed in a nonintegrable way, a symmetry breaking of
escaping rays occurs which can dramatically  modulate 
the outgoing wave even for small perturbations. The underlying 
mechanism does not occur in integrable models for which the ray 
families can be computed exactly and is described in this Letter
on the basis of canonical perturbation theory. Emission from 
deformed resonators is currently of
immense practical interest in the context of whispering-gallery
optical resonances of dielectric cavities and the approach outlined here
promises simple analytical characterisations in the important
case of small deformations.
\end{abstract}
\maketitle

Short-wavelength approximations have proved invaluable
in understanding and predicting the properties of 
optical resonators \cite{rev1,rev2}. They have in particular 
provided essential insight into the directional emission patterns 
that are observed when resonators are sufficiently asymmetric 
that ray dynamics display chaotic as well as regular 
behaviour \cite{ARC}, where features of wave chaos such as
chaos-assisted tunnelling \cite{CAT,CATNP,CATDL}, scarring \cite{SYL,scar} and 
dynamical localisation \cite{CATDL,DL}
have been shown to play a role. Surprisingly,
emission characteristics in the apparently benign limit of 
very weakly asymmetric resonators cannot readily
be described using this approach \cite{ecc1}.  The reason is 
that emission is essentially a tunnelling phenomenon whose
short-wavelength approximation demands that we 
extend the underlying ray families into complex space.
It has been established, however, that natural boundaries generically
intervene \cite{nb}, even in very slightly perturbed problems, which 
prevent analytic continuation as far as needed in the complex
domain \cite{me,RAT}. In other words, the ray-dynamical data demanded by 
short-wavelength approximations of emission simply does not exist.

In view of the current technological interest in  
optical resonators \cite{rev2}, simple analytical approaches to this problem 
are clearly of great value. We outline one such
solution in this Letter which works by substituting
for the nonexistent classical data approximate solutions
that can  be extended as far as required in the complex domain
while remaining sufficiently accurate for use in WKB 
approximations. A related approach has recently proved successful
in predicting tunnel splittings in near-integrable potentials \cite{us}.
Although restricted to relatively small deformations,
the solution is capable of predicting strongly directional 
emission patterns.

\begin{figure}[h]
\vspace*{-0.1cm}\hspace*{-0.1cm}
\psfig{figure=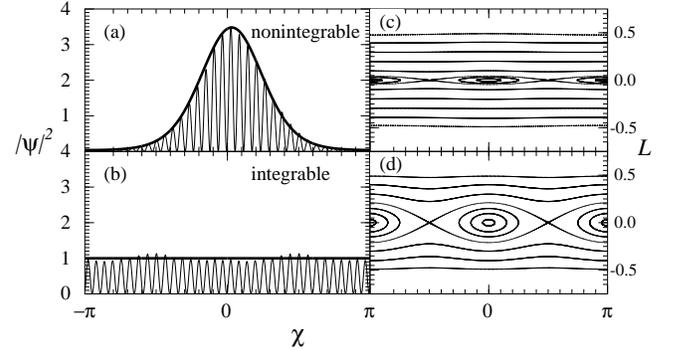,height=1.9in}
\vspace*{-0.7cm}
\caption{Emission patterns are shown for two perturbations
 of a resonance of
$V_0(r)=-(r^2-1)^2$, with the generic perturbation
$\eps x$ in (a) and the integrable perturbation 
$\eps x^2$ in (b).
In each case the thick line shows the envelope predicted by
first-order canonical perturbation theory and the thin line shows 
the angular dependence of $|\psi_{nm}|^2$, normalised to have unit average,
with $n=2$ and $m=17$ for $\eps=-1/40$ and $\hbar=1/40$.
The respective Poincar\'e plots, shown in (c) and (d),
are defined by setting $p_r=0$. Even though
the perturbations are of similar strength in each case, the primary 
islands near $L=0$  appear only at second order in the nonintegrable 
case and are much smaller, growing as $\eps$ rather than $\sqrt{\eps}$. 
}
\label{wellplot}
\end{figure}

A feature of the solution is that there is a dramatic
difference between the behaviours of \emph{integrable}
and \emph{nonintegrable} deformations. This is true
even of problems whose Poincar\'e plots look quite similar.
To illustrate this, emission patterns from two perturbations of a 
circular resonator, one integrable and the other nonintegrable,
are compared in Fig.~\ref{wellplot}. The details of these systems
are described later but here we simply point out that even 
though the Poincar\'e plot of the  nonintegrable problem
seems to deviate much less from the perfectly circular
limit, it has a dramatically more directional emission pattern.
The enormous qualitative difference between integrable
and nonintegrable systems, irrespective of any gross similarity 
of the real ray dynamics, indicates that simple 
geometrical characteristics of the deformation, such as eccentricity,
boundary curvature or the existence of particular island chains, 
do not transparently determine emission characteristics in this regime.

We note that a ``surprising'' directionality in the emission of 
weakly deformed resonators has previously been pointed out in
Ref.~\cite{ecc1} (see also \cite{ecc2,ecc3}), providing 
motivation for the current study. 
However, existing analysis of such directionality has been based 
on the structure of real phase space and the role played, for example,
by specific island chains. This seems distinct from the mechanism 
proposed here,  which 
applies at deformations so small that real ray dynamics might hardly be
affected. 
We will see that emission patterns can nevertheless be 
strongly modulated because
there is an essential topological
difference between the complexified dynamics of nonintegrable
and integrable perturbations, which essentially amounts to a
symmetry-breaking of the rays escaping to asymptopia.
These escaping rays are real
in the integrable case but are slightly complex for nonintegrable
perturbations and the resulting complex eikonal phase
strongly modulates the outgoing wave.

\begin{figure}[h]
\vspace*{-.0cm}\hspace*{.0cm}
\psfig{figure=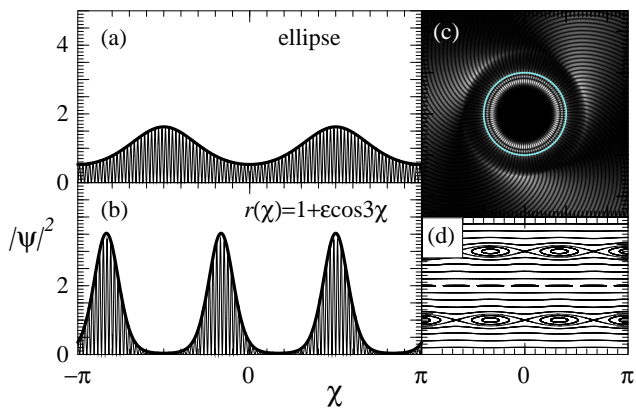,height=2.25in}
\vspace*{-.7cm}
\caption{Emission patterns are shown for two perturbations
of a circular resonator with refractive index
$n=2$. An elliptical deformation is shown in (a) and 
a generic perturbation $r= 1+\eps\cos3\chi$ is shown in (b).
In each case, $\eps=1/400$, where the elliptical perturbation
is $r\approx 1+\eps\cos2\chi$ at first order and the resonance has $m=50$
and $\Re(k)\approx 33.19$.
A two-dimensional illustration and a Poincar\'e-Birkhoff
plot are shown for the generic perturbation in (c) and (d) 
respectively. In (c), the intensity of the emitted wave has been
exaggerated to make it visible.}
\label{cavplot}
\end{figure}

As a simple model of a deformed resonator we consider 
the Schr\"odinger equation for a perturbation
 $V(\x)=V_0(r)+\eps V_1(\x)$ of a central potential well 
$V_0(r)$. This is analogous to a scalar optical problem
$
\nabla^2\psi(\x) + k^2 n^2(\x)\psi(\x) = 0
$
with refractive index and wave number defined by 
$k^2 n^2(\x)=2(E-V)/\hbar^2$ (assuming unit mass).
Of course, optical resonators have sharp material interfaces which should
be modelled by step potentials in the Schr\"odinger equation.
We present numerical evidence that a similar mechanism is at 
play for such systems in Fig.~\ref{cavplot}, where
strong directionality is found in emission from perturbations
of a circular resonator (Fig~\ref{cavplot}(a) and \ref{cavplot}(b))
which are so weak that there is little sign of nonintegrability 
in Poincar\'e plots (Fig.~(\ref{cavplot}(d)). We confine our
detailed analytical work in this Letter however to smooth (or more accurately,
analytic) potentials. The underlying idea can be set out more transparently 
this way and we will treat step potentials more properly in a 
future publication.
For similar reasons,  we will present details for the two-dimensional
case only and assert that it will be obvious how to adapt the 
discussion to three dimensions.

\vspace{-.3cm}

\begin{figure}[h]
\vspace*{0.cm}\hspace*{.0cm}
\psfig{figure=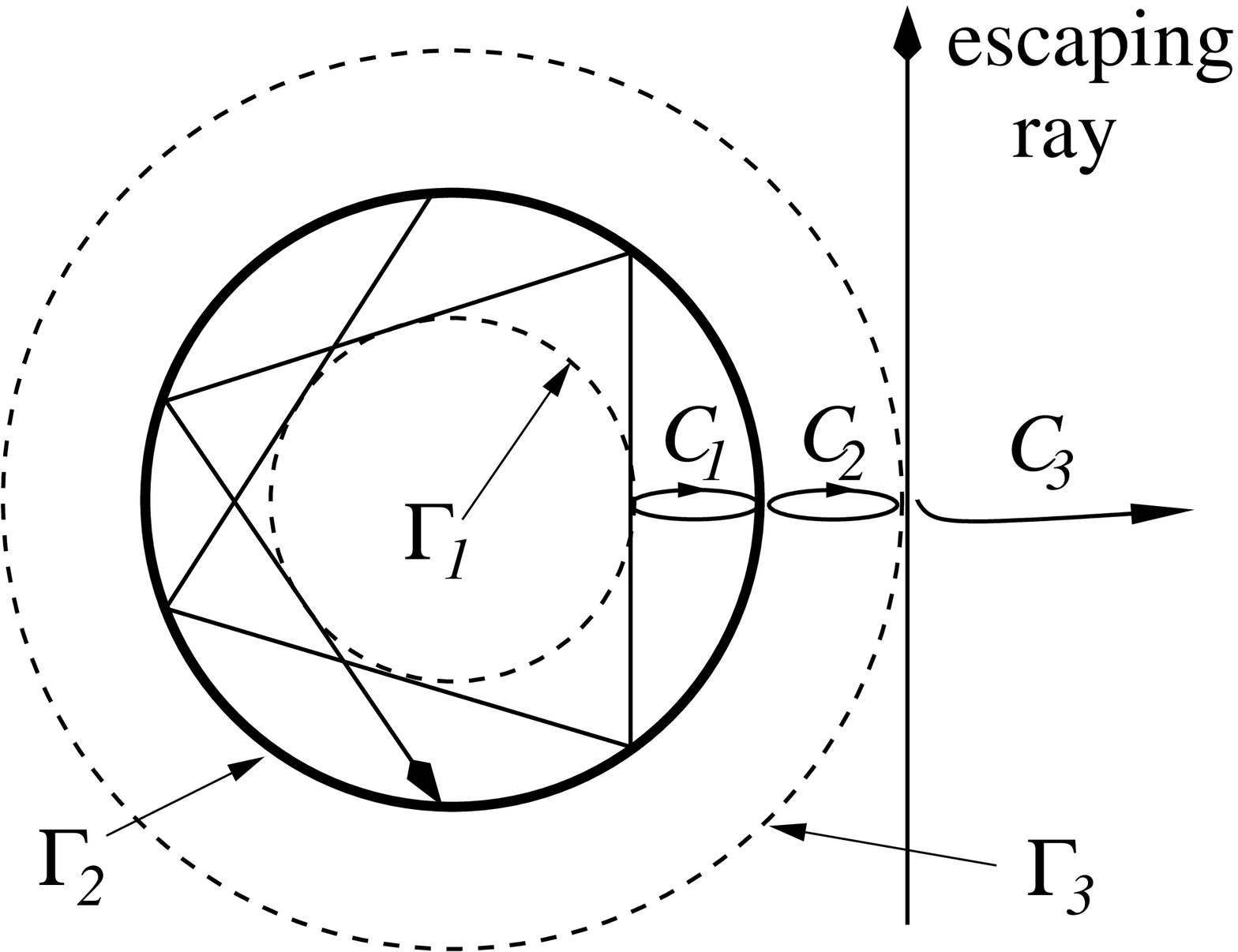,height=1.2in}\hspace*{-.2cm}
\psfig{figure=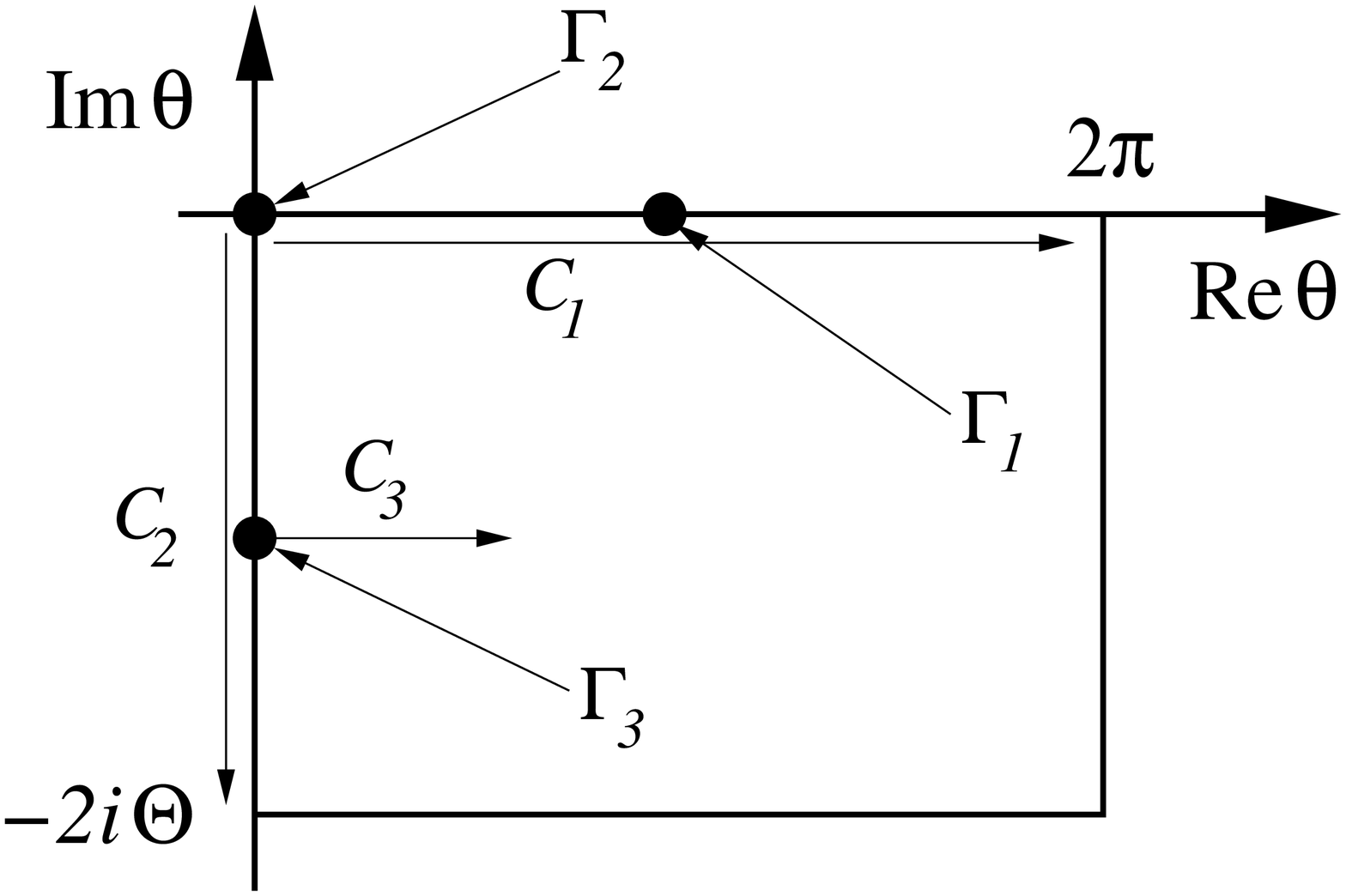,height=1.2in}
\caption{Real and complex rays 
are described in the complex angle plane as shown. $\C_1$, $\C_2$ and
$\C_3$ are caustics 
and  
$C_1$, $C_2$ and $C_3$ are contours in the complex $\theta$ plane
which define loops on these tori as shown at left.}
\label{circav}
\end{figure}

We consider a Gamow-Siegert
state $\psi_{nm}(\x)$, defined as a solution of the Schr\"odinger
equation for a complex energy $E=E_0-i\Gamma/2$ which satisfies 
radiating boundary conditions at infinity. Here $n$ and $m$ 
denote radial and azimuthal quantum numbers 
of the resonance in the limit of zero perturbation. 
In the short-wavelength
limit we can associate this state with an invariant torus of classical
rays moving in the plane configuration space in an annulus
bounded by inner and outer caustics $\C_1$ and $\C_2$,
as illustrated in Fig.~\ref{circav}.

Immediately outside $\C_2$ is a band of complex rays where 
$\psi_{nm}(\x)$ decays exponentially in the radial direction.
We argue below that this band  has an outer
caustic $\C_3$ where it joins with a family of rays escaping 
to infinity (see Fig.~\ref{circav}). It is these escaping rays which
determine the emission pattern. The wavefunction outside $\C_3$
has an eikonal form 
\begin{equation}\label{WKB}
\psi_{nm}(\x) \approx A(\x)\e^{iS(\x)/\hbar}
\end{equation}
where $S(\x)$ is the action function of the escaping rays.
We expand the action in the form
\[
S(\x) = S_0(\x) + \eps S_1(\x) +\cdots,
\]
where  $S_1(\x)$ is determined explicitly below using 
canonical perturbation theory. For integrable problems 
the escaping rays are necessarily real \cite{real} and so 
therefore is $\nabla S$. This implies that any angular variation
in the magnitude of the emitted wave 
derives from the amplitude $A(\x)$ and is slight if $\eps$ is 
small. If the escaping rays are complex on the other hand,
as we will argue they are in the generic
nonintegrable case, then  the emission
pattern is dominated by the imaginary part of the action
\[
|\psi_{nm}(\x)|^2 \propto  \e^{-2\eps\Im S_1(\x)/\hbar}
\]
at leading order. In this case a perturbation of the order 
of the small parameter $\hbar$ suffices to dramatically 
alter the envelope of the outgoing wave. We now set out 
the details of this calculation.

We let $(I,\theta)$ and $(L,\phi)$ respectively
denote the action-angle pairs of variables for the radial and azimuthal
degrees of freedom in the unperturbed limit and let
$\omega$ and $\Omega$ denote the corresponding
frequencies. Because the actions are fixed by the
torus quantisation rules $I=(n+\ha)\hbar$ and $L=m\hbar$
at leading order, we will suppress
the actions notationally in much of what follows, writing
$r(\theta)$ instead of $r(\theta,I)$ for example. We denote
by $(r,\chi)$ the standard polar coordinates in the plane and point out
that $\chi$ and $\phi$ are distinct but related by a shift
%\cite{Goldstein}
\[
\chi = \phi + \del(\theta),
\]
where
\[
\del(\theta) = \frac{1}{\omega} \int_0^\theta
\left(\frac{L}{r^2(\theta')}-\Omega\right)\d\theta'.
\]
We adapt the convention that $\theta=0$ on the outer caustic
$\C_2$ of the real torus (Fig.~\ref{circav}).
To characterise emission we must construct
the complex rays in the classically forbidden region outside 
$\C_2$. We first describe how this is achieved 
in the unperturbed limit. 

Starting
on the caustic $\C_2$ where  $\theta=0$, and letting $\theta$ move down the 
imaginary axis (the rays which define decaying WKB solutions
are in the lower-half plane), we find that the 
radial coordinate
$r(\theta)$ evolves periodically. Denote the 
imaginary period  by $2i\Theta$, so that 
\begin{equation}\label{imper}
r(\theta+2i\Theta) = r(\theta).
\end{equation}
As angles $(\theta,\chi)$ respectively range over the imaginary and 
real axes, a two-dimensional complex extension of the real torus
is swept out 
which itself has the topology of a torus and 
on which radial momentum $p_r$ is 
imaginary and $r$ is real. This second torus describes a band of evanescent 
decay immediately outside the resonator. It has an inner caustic $\Gamma_2$,
where it touches the real torus, and an outer caustic $\C_3$.
The caustic $\C_3$ corresponds to $\theta=-i\Theta$. If $\theta$ evolves
horizontally from  $-i\Theta$ (contour $C_3$ in Fig.~\ref{circav}) 
then we describe a family of real orbits escaping to infinity.

The crucial feature in this description (of the unperturbed case) 
is that the orbits escaping to infinity are \emph{real}. This 
is related in the following way to the  fact that  the
radial coordinate and other phase space functions are \emph{biperiodic}
functions of $\theta$ --- in addition to the usual real period
$r(\theta+2\pi) = r(\theta)$ 
there is an imaginary period expressed in (\ref{imper}).
We have stated that the escaping rays are launched from the contour 
$C_3$ in the complex plane (see Fig.~\ref{circav}).
They are real if they coincide with their complex conjugates,
which are found along the conjugate contour $C_3^*$. Since
$C_3^*=C_3+2i\Theta$, the rays are therefore real if
there is an imaginary period, as claimed.
We will now show that if the rays are approximated using canonical 
perturbation theory, this second period is destroyed at first 
order and the rays therefore become complex.

We use a type-two  generating function
$F_2(\theta,\phi,\Ibar,\Lbar)
=\theta \Ibar + \phi \Lbar + \eps G(\theta,\phi,\Ibar,\Lbar)$ to generate
the transformation to action-angle variables 
$(\bar{\theta},\bar{\phi},\Ibar,\Lbar)$ for the perturbed system.
At first order,
$G(\theta,\phi,\Ibar,\Lbar)$ provides 
the leading correction to the phase of 
$\psi_{nm}(\x)$  and is a solution of 
\begin{equation}\label{pt1}
\omega\dydxv{G}{\theta} + \Omega\dydxv{G}{\phi} = -\V(\theta,\phi),
\end{equation}
where $\V(\theta,\phi)$ is the oscillating part of $V_1(\x)$,
expressed in action-angle variables \cite{us}. 
As in the unperturbed problem, 
$(\Ibar,\Lbar)$ are fixed by torus quantisation conditions 
and are suppressed notationally from now on.

The perturbing potential  $\V(\theta,\phi)$ inherits a biperiodic 
structure from the radial degree of freedom as follows. We first note 
that
\[
\del(\theta+i\Theta) = \del(\theta)-i\ddel,
\]
where
\[
\ddel = \int_0^{\Theta/\omega}
\left(\Omega-\frac{L}{r^2(i\omega\tau)}\right)\d\tau
\]
is real. The potential is periodic with respect to
$(\theta,\chi)\to(\theta+2i\Theta,\chi)$, which, in terms
of dynamical angles, gives 
\[
\V(\theta+2i\Theta,\phi+2i\ddel)=\V(\theta,\phi).
\]
Let the perturbing potential have the form
\[
\V(r,\chi) = \sum_k U_k(r)\e^{ik\chi} = \sum_k \v_k(\theta)\e^{ik\phi}
\]
where $\v_k(\theta)= U_k(r(\theta))\e^{ik\del(\theta)}$. 
We substitute
\[G(\theta,\phi)=\sum_k g_k(\theta)\e^{ik\phi}\] 
in (\ref{pt1}) and solve the resulting equation for $g_k(\theta)$
to get
\[
g_k(\theta) = \e^{-ik\Omega\theta/\omega}g_k(0)
-\frac{\e^{-ik\Omega\theta/\omega}}{\omega} \int_0^\theta 
\e^{ik\Omega\theta'/\omega}\v_k(\theta')\d\theta'.
\]
The integration constant  $g_k(0)$ is fixed by imposing the 
period $g_k(\theta+2\pi)=g_k(\theta)$ on the real axis. The resulting
generating function is single-valued on the real axis, where it
gives the real torus between caustics $\C_1$ and $\C_2$.

Emission patterns are dominated by the imaginary
part of $G(\theta,\phi)$ along $C_3$. In light of
$C_3^*$ = $C_3+2i\Theta$, this can be written $\Im\, G(\theta,\phi)=
\Delta G(\theta,\phi)/(2i)$, where
\[
\Delta G(\theta,\phi) = G(\theta,\phi)-G(\theta+2i\Theta,\phi+2i\ddel).
\]
It can be shown that 
\[\Delta G(\theta,\phi)= i\a_0 + i\A(\theta,\phi),
\]
where 
\[
\a_0 = \frac{1}{i\omega}\int_0^{2i\Theta} \v_0(\theta)\d\theta
= \int_0^{2\Theta/\omega} \v_0\left(i\omega\tau\right)\d\tau
\]
is a real constant and
\[
\A(\theta,\phi) =
\sum_{k\neq 0} \frac{\e^{ik(\phi-\Omega\theta/\omega)}}
{1-\e^{2\pi i k\Omega/\omega}}\;
\frac{1}{i\omega}
\oint_C \e^{ik\Omega\theta'/\omega}\v_k(\theta')\d\theta'.
\]
The contour $C$ is a clockwise circuit of 
the rectangle at right in Fig.~\ref{circav} (a 
derivation of an analogous result in the context of 
double-well splittings can be found in \cite{us}).

It is convenient to express the result using the angles
\begin{eqnarray*}
\alpha(\theta,\phi) &=& \phi+i\Phi-\frac{\Omega}{\omega}(\theta+i\Theta),\\
\beta(\theta) &=& \frac{\Omega}{\omega}(\theta+i\Theta)+\del(\theta+i\Theta)= 
\frac{1}{\omega} \int_{-i\Theta}^\theta \frac{L}{r^2(\theta')}\d\theta'.
\end{eqnarray*}
Note that $\alpha$ is real and constant along the escaping rays of 
the unperturbed problem and is the value of the polar angle $\chi$
at which a given ray touches the caustic $\C_3$. We also have 
$\chi=\alpha+\beta$, so $\beta$ measures the angular deflection
of the escaping ray as it moves away from $\C_3$. Then
\[
\A(\theta,\phi) = \sum_{k\neq 0} \a_k \e^{ik\alpha}
\]
where
\[\a_{k\neq 0} = \frac{1}
{1-\e^{2\pi i k\Omega/\omega}} \frac{1}{i\omega}
\oint_C \e^{ik\beta(\theta')}U_k(r(\theta'))\d\theta'.
\]
The main conclusion of this Letter is that 
the asymptotic intensity of a WKB mode (\ref{WKB}) 
is modulated by a function
\begin{equation}\label{mod}
\M\left(\alpha\right)=\e^{-\eps \A(\alpha)/\hbar}
\end{equation}  
which can vary 
strongly with $\chi$ even for 
$\eps=O(\hbar)$.

We illustrate this result numerically for the potential 
$V_0(r)=-(r^2-1)^2$. Although this potential does not vanish at infinity,
it defines Gamow-Siegert states entirely analogous to those
of asymptotically free problems and is used because solutions
are easily obtained for it using complex rotation in a 
harmonic-oscillator basis. None of the calculations above
are substantially affected by this choice. We consider two perturbations,
$V_1(\x)=x$, which behaves generically, and 
$V_1(\x)=x^2$, for which the total potential can be 
shown to be separable in elliptic coordinates and which therefore
provides us with an example of an integrable perturbation.
The respective angular distributions of the outgoing intensity, normalised
to have unit average, are shown 
in Figs.~\ref{wellplot}(a) and \ref{wellplot}(b).
The heavy curves show the modulation function 
(\ref{mod}). Note that for the integrable perturbation we necessarily 
find that $\A(\alpha)=0$ and the small modulation in the numerical solution
is a result of corrections to the amplitude $A(\x)$, which 
have not been taken into account. In the nonintegrable case, the 
modulation argument can be written $\A(\alpha) = 2\a_1\cos\alpha$ 
and provides a good 
description of the numerical emission pattern (Fig.~\ref{wellplot}(a)).

A similar modulation is seen in the numerical results for a deformed 
cavity shown in Fig.~\ref{cavplot}. The dielectric ellipse is a special case
--- it is not separable but the real and complex
rays can nonetheless be calculated analytically. The envelope shown
is $\M(\alpha) = \exp[2{\eps k (n^2-1)\sqrt{n^2 p^2-1}\cos2\alpha}]$,
where $p=m/k$, whose derivation will be described in a future 
publication. The envelope for the generically nonintegrable perturbation
in (b) has not been calculated analytically but has been found numerically
to 
be well described by a function of the form 
$\M(\alpha)=\exp[{\eps k a(p)\cos3\alpha}]$,
with a typical case shown in the figure. This is 
consistent with the mechanism set out in this Letter.

In conclusion, canonical perturbation theory married with complex
WKB approximation can successfully describe strongly directional 
emission from weakly deformed resonators. The underlying mechanism 
is not transparently related to the structure of real phase space
and there is a stark qualitative difference between integrable and 
nonintegrable systems, even when the signatures of nonintegrablilty
are slight in real phase space \cite{chaosremark}.
Despite this sensitivity to dynamical detail,
simple analytical formulas can be given for the observed output.
The theory has been presented for smooth potentials but a similar
mechanism is expected to govern sharp cavity problems and should
provide a useful analytical tool to understand the evanescent field
outside optical resonators, which is important to applications such
as lasers and sensors \cite{rev2}.

%\end{bibliography}


\begin{references}
%\begin{bibliography}
\bibitem{rev1} H. G. L  Schwefel et al,  in 
\emph{Optical Microcavities} ed. K. Vahala (World Scientific, 2004).
\bibitem{rev2} 
A. B. Matsko and V. S. Ilchenko, IEEE Journal of 
Selected Topics in Quantum Electronics \textbf{12} 3 (2006); 
V. S. Ilchenko and A. B. Matskolm, IEEE Journal of 
Selected Topics in Quantum Electronics \textbf{12} 15 (2006). 
\bibitem{ARC}J. U. N\"ockel, and A. D. Stone, Nature {\bf 385}, 45 (1997);
C. Gmachl et al, Science {\bf 280}, 1556 (1998).
\bibitem{CAT} O. Bohigas, S. Tomsovic and D. Ullmo, Phys. Rep. {\bf 223}, 
43 (1993);  S. Tomsovic and D. Ullmo, Phys. Rev. E {\bf 50}, 145 (1994);
F. Leyvraz and D. Ullmo, J. Phys. A {\bf 29}, 2529 (1996).
\bibitem{CATNP} V. A. Podolskiy and E. E. Narimanov,  Opt. Lett. {\bf 30}, 
474 (2005).
\bibitem{CATDL} E. E. Narimanov and V. A. Podolskiy, IEEE Journal of 
Selected Topics in Quantum Electronics \textbf{12} 40 (2006);
\bibitem{SYL} S.-Y. Lee et al, Phys. Rev. Lett. {\bf 93}, 164102 (2004);  
S.-Y. Lee et al, Phys. Rev. A {\bf 72}, 061801(R) (2005).
\bibitem{scar} C. Gmachl et al, Opt. Lett. {\bf 27}, 
824 (2002).
\bibitem{DL}  V. A. Podolskiy et al, Proc. Natl. Acad. Sci. USA {\bf 101},
10498, (2004); W. Fang et al, Optics Express {\bf 13}, 5641 (2005).
\bibitem{ecc1}S. Lacey et al, Phys. Rev. Lett. {\bf 91}, 033902 (2003).
\bibitem{nb} I. C. Percival and J. M. Greene, Physica D {\bf 3}, 530 (1981);
 I. C. Percival, Physica D {\bf 6}, 67 (1982); L. Billi, G. Turchetti and
R. Xie, Phys. Rev. Lett. {\bf 71}, 2513 (1993).
\bibitem{me} S.C. Creagh, in {\it Tunneling in Complex Systems},
S. Tomsovic (ed.), World Scientific, Singapore, (1998).
\bibitem{RAT} O. Brodier, P. Schlagheck and D. Ullmo, 
Phys. Rev. Lett. {\bf 87} 064101 (2001); Ann. Phys. {\bf 300}, 88 (2002).
\bibitem{us} G. C. Smith and S. C. Creagh, J. Phys. A {\bf 39}, 8283 (2006).
\bibitem{ecc2} H. G. L. Schwefel et al, J. Opt. Soc. Am. B {\bf 21}, 
923 (2004).
\bibitem{ecc3} S-K Kim et al, Appl. Phys. Lett. {\bf 84}, 863 (2004).
\bibitem{real} Integrable systems are characterised by the existence
of a global invariant which is real-valued on real phase space.
The escaping family of rays is determined by fixing this invariant, 
along with energy, and is seen therefore to be real. (The very small 
imaginary part of the energy $E=E_0-i\Gamma/2$ leads to some modulation 
along individual rays but does not affect angular 
dependence and is ignored.)

% Complex
%escaping rays therefore imply the absence of a real global invariant,
%which in turn implies nonintegrability.
%\bibitem{comment} The imaginary period in the complex $\theta$ plane
%is simply a manifestation of the the under-barrier 
%instanton orbit in the one-dimensional effective 
%radial potential.
%\bibitem{Goldstein} H. Goldstein. C. P. Poole and J. L. Safko, 
%{\it Classical Mechanics}, 3rd edition (Addison Wesley, 2002).
% The analogous situation for a nonseparable perturbation
%is less obvious, however.
\bibitem{chaosremark} {The total area of chaotic layers in 
Fig.~\ref{wellplot}(a) is many orders of magnitude smaller than $\hbar$, 
for example.} 
\end{references}
\end{document}